\documentclass[conference]{IEEEtran}
\IEEEoverridecommandlockouts
\usepackage{cite}
\usepackage{amsmath,amssymb,amsfonts}
\usepackage{graphicx}
\usepackage{textcomp}
\usepackage{multirow}
\usepackage[multiple]{footmisc}
\usepackage{xcolor}
\usepackage{booktabs}
\usepackage{eso-pic}
\usepackage{algorithm}
\usepackage{algpseudocode}
\usepackage{balance}
\usepackage{threeparttable}
\usepackage{tabularx}
\def\BibTeX{{\rm B\kern-.05em{\sc i\kern-.025em b}\kern-.08em
    T\kern-.1667em\lower.7ex\hbox{E}\kern-.125emX}}
\newcolumntype{P}[1]{>{\centering\arraybackslash}p{#1}}

\usepackage[hyperfootnotes=false]{hyperref} 

\newcommand\copyrighttext{%
  \footnotesize \textcopyright 2022 IEEE. Personal use of this material is permitted.
  Permission from IEEE must be obtained for all other uses, in any current or future
  media, including reprinting/republishing this material for advertising or promotional
  purposes, creating new collective works, for resale or redistribution to servers or
  lists, or reuse of any copyrighted component of this work in other works.
  DOI: \href{https://doi.org/10.1109/BigData55660.2022.10021129}{https://doi.org/10.1109/BigData55660.2022.10021129}}
\newcommand\copyrightnotice{%
\AddToShipoutPicture*{%
\put(45,30){%
\centering
\fbox{\parbox{\dimexpr\textwidth-\fboxsep-\fboxrule\relax}{\copyrighttext}}
}}}

\DeclareMathAlphabet{\altmathcal}{OMS}{cmsy}{m}{n}
    
\begin{document}

\title{Probabilistic Time Series Forecasting for Adaptive Monitoring in Edge Computing Environments}

\author{
\IEEEauthorblockN{
Dominik Scheinert\IEEEauthorrefmark{1},
Babak Sistani Zadeh Aghdam\IEEEauthorrefmark{1},
Soeren Becker\IEEEauthorrefmark{1},
Odej Kao\IEEEauthorrefmark{1},
and Lauritz Thamsen\IEEEauthorrefmark{2}
}
\IEEEauthorblockA{
\IEEEauthorrefmark{1}
Technische Universit{\"a}t Berlin, Germany, \{firstname.lastname\}@tu-berlin.de}
\IEEEauthorblockA{
\IEEEauthorrefmark{2}
University of Glasgow, United Kingdom, lauritz.thamsen@glasgow.ac.uk}
}

\maketitle
\copyrightnotice

\begin{abstract}
With increasingly more computation being shifted to the edge of the network, monitoring of critical infrastructures, such as intermediate processing nodes in autonomous driving, is further complicated due to the typically resource-constrained environments. 
In order to reduce the resource overhead on the network link imposed by monitoring, various methods have been discussed that either follow a filtering approach for data-emitting devices or conduct dynamic sampling based on employed prediction models.
Still, existing methods are mainly requiring adaptive monitoring on edge devices, which demands device reconfigurations, utilizes additional resources, and limits the sophistication of employed models.

In this paper, we propose a sampling-based and cloud-located approach that internally utilizes probabilistic forecasts and hence provides means of quantifying model uncertainties, which can be used for contextualized adaptations of sampling frequencies and consequently relieves constrained network resources.
We evaluate our prototype implementation for the monitoring pipeline on a publicly available streaming dataset and demonstrate its positive impact on resource efficiency in a method comparison.
\end{abstract}


\section{Introduction}
\label{sec:introduction}

The amount of devices and sensors deployed in the Internet of Things (IoT) is increasing steadily. At the same time, this coincides with a rapid increase of generated data by the employed devices.
Traditional cloud computing architectures encounter problems when trying to cope with this increasing  scale, as new use cases, e.g. smart cities and manufacturing, digital health care, or autonomous driving pose considerable challenges to the underlying infrastructure. Especially in the aforementioned domains, the amount of collected and transferred data to the cloud adds an additional burden on possibly unreliable network connections and renders latency-bounded and bandwidth-intensive applications infeasible \cite{yu2017survey, edgepier}.

In order to unburden the network and improve overall communication efficiency, the edge computing paradigm has been gaining momentum in the past years. By shifting computing capabilities closer to the actual data sources at the edge of the network, such environments enable the processing of data on edge devices in highly distributed architectures \cite{cloudcontinuum, efficientprofiling}.
The employed edge devices are typically resource-constrained and remotely located. Thus, they can i.e. easily be overloaded and are vulnerable to damage or theft.
Since these critical infrastructures can have a decisive impact on everyday life, continuous monitoring is required in order to detect problems early on and assure the expected functionality \cite{aiopsedge}.
However, constantly transmitting all the monitoring data, especially with high frequency, consumes noticeable network bandwidth \cite{venkateswaran2020ream} and can thus in turn aggravate network congestions or even service interrupts \cite{lou2020data,venkateswaran2020ream,zhai2020iot,ma2021one}.
Therefore, the transmission rate of monitoring data is often reduced \cite{venkateswaran2020ream}, i.e. by adaptively adjusting the monitoring rate \cite{venkateswaran2020ream,trihinas2015adam,TrihinasPD17}.
Several approaches \cite{tata2017optimization,FathyBT18} embed and employ the adaptive functionality directly on the edge devices, and although this can yield promising results, it also results in further  processing load on already resource-constrained nodes.

Hence, in this paper, we are proposing an adaptive monitoring approach that is deployed on cloud nodes and chooses the monitoring frequency based on forecasting future monitoring metrics. Exploiting probabilistic forecasting models to decide whether to fetch monitoring data from edge nodes or to rely on the forecasting output, we aim to reduce monitoring traffic and at the same time still allow for optimizing and automating operations based on accurate monitoring metrics. Accordingly, our approach considers the variability of the data over time and retrains periodically in order to prevent the drifting of the forecasting model \cite{cavalcante2016fedd}.

\emph{Contributions}. The contributions of this paper are:
\begin{itemize}
    \item A design for a system that minimizes data transmission rates in resource-constrained environments via sampling-based adaptive monitoring.
    The internally used probabilistic forecasting method allows for the assessment of predictions and hence conditional sampling.
    \item A prototypical implementation of our adaptive monitoring routine which follows the outlined principles of our proposed system design and is therefore representative.
    \item An evaluation of our implementation on a publicly available streaming dataset and comparison to a related method.
    We demonstrate the effective reduction of data transmission rates while retaining accurate metric data estimates, and discuss the implications of our findings.
\end{itemize}

\emph{Outline}. 
\autoref{sec:related_work} discusses the related work.
\autoref{sec:idea} elaborates on the idea and proposes a system for sampling-based and adaptive monitoring, whereas \autoref{sec:approach} concretizes on the modeling approach for probabilistic forecasting of metrics.
\autoref{sec:results} presents the preliminary results of our comparison with a related method, and a discussion of general requirements of our approach.
\autoref{sec:conclusion} concludes the paper.
\section{Related Work}
\label{sec:related_work}

This section discusses various related methods for adaptive monitoring as well as their differences from our method.

\subsection{Adaptive Filtering}

Solely emitting data values when they significantly differ from past data values is a strategy implementable on the device level and referred to as adaptive filtering.
JCatascopia~\cite{TrihinasPD14} is a monitoring framework that adjusts the filtering range in regard to a user-defined threshold, where the threshold defines the percentage of the data that should be filtered.
The ATOM framework~\cite{DuL15} follows a similar approach and additionally sends the median values of a subset of metrics when the values of these metrics did not change more than a threshold in a certain period of time.
Again another data filtering system~\cite{KimJK17} conducts data filtering when previously observed patterns in the data are maintained, which is achieved by training a classifier on past data and using it on new data. 

In contrast, our proposed framework only assumes accessible metric endpoints on the source nodes, which makes our approach agnostic against the concrete set of metrics that shall be modeled.
Running the modeling solution on sink nodes further allows for the employment of more sophisticated methods due to non-existent battery constraints.

\subsection{Adaptive Sampling}

Adjusting the interval of sampling target devices in a dynamic manner, based on observed data characteristics, is a strategy called adaptive sampling.
PayLess~\cite{ChowdhuryBAB14} is an adaptive monitoring framework that adjusts the sampling frequency by an operation with a constant definable value based on the difference between the current and the previous monitoring data and a predefined threshold.
Another work~\cite{MengL13} is based on the violation-likelihood detection method: The likelihood of not detecting a violation between two successive data points is calculated, which is used as an indicator, together with a user-defined threshold, for either establishing a fixed interval or conducting more frequent monitoring.
FAST~\cite{FanX14} is a framework that evaluates the need of adapting the sampling interval at each time step and adjusts the sampling frequency based on the error between a prior and a posterior estimation.
With EASA~\cite{SrbinovskiMMPP16}, the authors propose an energy-aware method that attempts to determine the optimal sampling frequency and takes the battery level of target IoT devices into consideration.

With our approach using probabilistic forecasts and a robust model update strategy, we tackle limitations such as the negligence of evolutionary data streams, missing or insufficient model update strategies, and non-existent uncertainty handling.

\subsection{Hybrid Algorithms}

This category encompasses methods that either combine adaptive filtering and adaptive sampling, or optimize not only the amount of transmitted data.
ADMin~\cite{TrihinasPD17} is a framework that aims to reduce the data which is produced in a network and also reduce energy consumption by devices. 
Data is published over the network solely when a shift is detected in the data stream.
The same authors also propose the AdaM framework~\cite{TrihinasPD21}, which measures the streaming data variability and evolution alongside employing two algorithms for adaptive sampling and adaptive filtering in order to reduce the monitoring data disseminated throughout the network.
With the AM-DR framework~\cite{FathyBT18}, the authors attempt to reduce the data transmission between the sink and sensor nodes by  predicting readings at both the source and sink nodes and transmitting sensor data when the difference between predicted and observed values exceeds a predefined threshold.
The SETAR framework~\cite{ArbiDS17} employs a forecasting model for both the data aggregation layer and the source node, and in case the forecasted metrics differ more than a threshold from the actual values, the transmission of the actual values is triggered.
In other works~\cite{LealLCF14,RazaCMPP15}, the authors propose to fit a linear model on the recent sensed values and only send the updates of the model parameters to the sink node.
A new model is computed and distributed among nodes if the predicted values continue to deviate from the actual sensed values.
Efficient sampling and hence reduced data transmission rates are also the result of a distributed Active Learning framework~\cite{NedelkoskiTVK19} for IoT applications deployed on multi-layer infrastructures.

The majority of hybrid algorithms require hardware or software modifications for both the source node and sink node, which makes their application challenging and less straightforward.
With our approach, we are relieving the IoT devices of interest and hence simplify the operation.

\section{Proposed System}
\label{sec:idea}

\begin{figure}[h]
    \centering
    \includegraphics[width=.8\columnwidth]{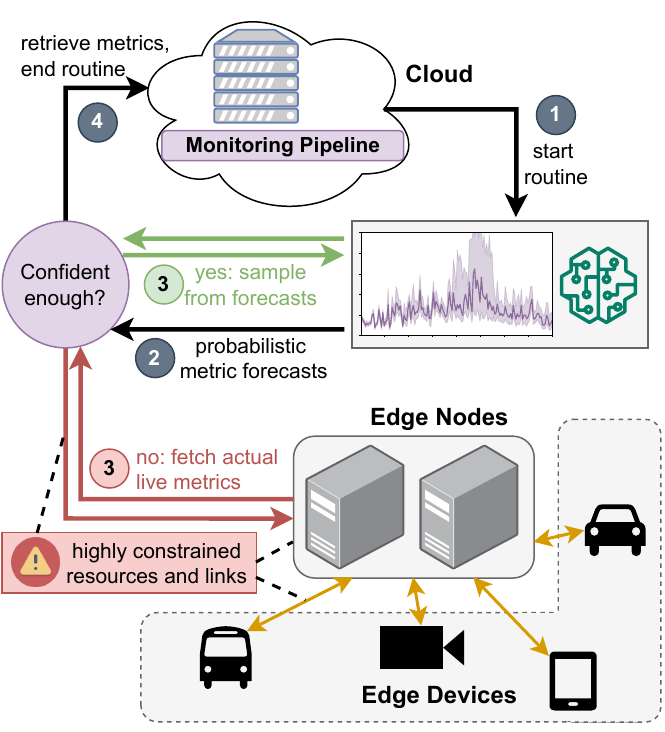}
    \caption{Overview of the envisioned system.
    Limited computing resources and network capacity characterize the edge environment.
    Hence, we reduce the overall network usage by conditionally sampling from either a target node or a model distribution, based on the models' confidence in its predictions.}
    \label{fig:system_overview}
\end{figure}

This section elaborates on our envisioned system for adaptive monitoring in resource-constrained environments via a sampling-based approach. 
It is further illustrated in~\autoref{fig:system_overview}.

\subsection{Probabilistic Adaptive Sampling}

Monitoring of services or devices in resource-constrained environments comes with its own challenges.
In order to reduce the additional bandwidth usage caused by monitoring, methods have been proposed in the past that seek a minimization of transferred information.
Yet, most of them require changes to the respective source node and/or sink node, or provide insufficient means of dealing with imprecise predictions and the evolution of data streams over time. 
For applicability in real-world scenarios, it is therefore desirable to design an approach that shifts control back to the respective sink node(s), requires no changes on the device level and is hence fairly agnostic, and employs a sophisticated strategy for dealing with model predictions and variability in data streams. 
Consequently, we demand a sampling-based approach that utilizes probabilistic forecasting to realize adaptive monitoring.

\subsection{Envisioned System}

Assuming that target devices expose relevant metrics by factory default or this functionality can be retrofitted with manageable effort, the amount of transmitted data can be reduced using a sampling-based approach, i.e., by employing a prediction model on a sink node. 
Once sufficient data has been collected via initial frequent sampling, a probabilistic prediction model can be trained and used for predicting future values together with a notion of prediction uncertainty.
From there on, at any point in time, the range of possible values predicted by the model is evaluated.
In case of significant model uncertainty, actual metric data is sampled from the source node, whereas otherwise, we sample from the aforementioned value range of the model.
With this envisioned system, the overall data transmission rate can be reduced and a conscious strategy for dealing with uncertainties is enabled.

\section{Resource-Efficient Adaptive Monitoring}
\label{sec:approach}

\begin{figure}[h]
    \centering
    \includegraphics[width=.8\columnwidth]{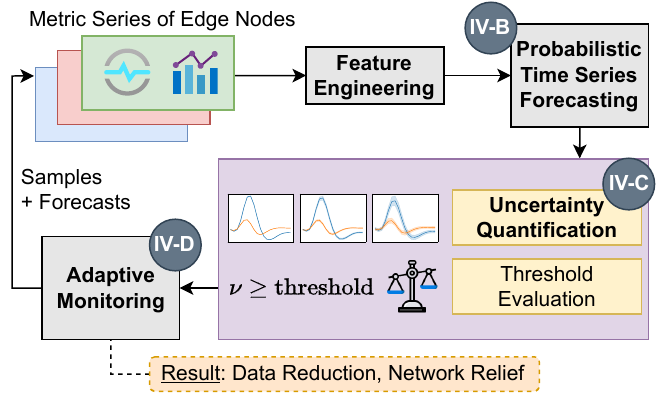}
    \caption{The proposed pipeline.
    Time series data of actual metrics are processed, enriched, and used as forecasting model input.
    The model output is then further analyzed and evaluated with regard to adaptive monitoring.}
    \label{fig:approach_pipeline}
\end{figure}

This section presents our approach to adaptive monitoring in edge computing environments using probabilistic time series forecasts. 
The approach is generally sketched in~\autoref{fig:approach_pipeline} and explained in more detail in the following paragraphs.

\subsection{Preliminaries}

When collected over time, metric data can provide an abstract representation of the state of each system component.
As in our previous work~\cite{scheinertCloudIntelligence}, we define metric data as multivariate time series, i.e. a temporally ordered sequence of vectors $S = ({S}_t \in \mathbb{R}^d : t=1,2,\ldots, T)$, where $d$ is the number of metrics and $T$ defines the last sample time stamp.
For $S^{a}_{b}=(S_a, S_{a+1}, \ldots, S_b)$, we denote indices $a$ and $b$ with $a \leq b$ and $0 \leq a,b \leq T$ as time series boundaries in order to slice a given series $S^{0}_{T}$ and acquire a subseries $S^{a}_{b}$.
Additionally, we use the notion $S(i)$ to refer to a particular metric dimension $i$, with $1 \leq i \leq d$. 
With $\hat{S}$, we furthermore refer to a forecasted multivariate time series.

\subsection{Probabilistic Time Series Forecasting}

The problem with commonly employed deterministic forecasting approaches is that an indication of the certainty of model outputs is missing. 
In our adaptive sampling setting, this hinders the quality assessment of model forecasts and would only allow for an evaluation in retrospect. 
Conveniently, we can make use of probabilistic forecasts to tackle this limitation. 
Here, the underlying idea is that of \emph{quantile regression}, where the loss is formally defined and computed as
\begin{equation}
L_{\rho}(y, \hat{y})=\rho\cdot f(y-\hat{y})+(1-\rho)\cdot f(\hat{y}-y),
\end{equation}
with $\rho \in (0,1)$ being the quantile, $f(x) = max(0,x)$ a smoothing function of predicted values, $y$ the ground truth, and $\hat{y}$ the corresponding predicted sample.
Consequently, in the case of our previously defined multivariate time series, the total model loss is then calculated across time ($T$), quantiles ($P$), and the user-defined prediction horizon $K$ as:
\begin{equation}
\sum_t^T \sum_{\rho}^P L_{\rho}\left(S^t_{t+K}, \hat{S}^t_{t+K}\right).
\end{equation}
By training a model with this loss function and adapting the model parameters accordingly, we can assess the certainty of model predictions later, which is imperative for our approach to adaptive monitoring.

\subsection{Uncertainty Quantification}

We attempt to train a model on all metrics of a target system, which results in a multivariate time series where we can possibly exploit correlations between individual metrics.
Next, for each individual metric, we regulate the corresponding sampling frequency, which demands a suitable criterion.
In order to quantify the uncertainty of each metric, we utilize the standard deviation as a criterion to quantify the variability of predicted samples across all quantiles. 
If the forecasted samples have variance more than a predefined threshold, intuitively, we consider the model outputs to be uncertain due to the wide range of possible values.
If we sample $N$ values from the model's learned distribution, then the uncertainty quantification process can be formulated as below:

\begin{equation}
\sigma_{k}=\sqrt{\frac{\sum_{i=1}^N\left(s_{i}-\mu_{k}\right)^{2}}{N}},\  \nu=\frac{\sum_{k=1}^K\sigma_{k}}{K}
\end{equation}

Here, $\sigma_{k}$ is the standard deviation of the forecasting samples at time $k$, $s_i = \hat{S}^k_k$ is the $i$-th sample, and $\mu_{k}$ is the average of all $N$ samples at time step $k$. 
At the next step, the average of all standard deviations of all time steps in the forecasting windows length, i.e. $K$, is calculated as $\nu$. 
Finally, the value of $\nu$ is compared with the defined threshold, which triggers a new sampling routine based on the outcome of this evaluation.
Evidently, this uncertainty quantification can also be used as a trigger for model retrainings, e.g., if it is observed that defined thresholds are violated more frequently than before.

\subsection{Adaptive Monitoring}

With a strategy now in place for model training and conditional sampling, we can further elaborate on our overall routine for adaptive monitoring, for which we summarize the pseudocode in~\autoref{alg:pipeline}.
In the first step, all metrics are fetched from the respective target system for the duration of the defined input window length of the model. 
In the next step, the model predicts future metric values based on the given input for the duration of the defined forecasting horizon length. 
Afterward, as previously described, the uncertainty of each metric for the period of the forecasted time is calculated. 
If no metric with high uncertainty is found, then the next input of the model will be set to the recently forecasted series. 
Otherwise, our monitoring routine idles until the last forecasted time step with high certainty and then triggers the fetching of uncertain metrics. 
Subsequently, the fetched metrics and forecasted metrics are combined along the time dimension and used as the next model input.

\begin{algorithm}
\small
\caption{Pseudocode of adaptive monitoring routine}
\label{alg:pipeline}
\begin{algorithmic}
\State $N$ \Comment{forecasting horizon length}
\State $L$ \Comment{input window length}
\State $FS$ \Comment{forecasted series}
\State $FM$ \Comment{fetched metrics}
\State $UM$ \Comment{metrics with high uncertainty}
\State $IS \gets fetchAllMetrics(length = L)$ \Comment{input to model}
\State $LFT \gets 0$ \Comment{last forecasted time step}
\While{TRUE}
\State $FS \gets forecast(input = IS)$
\State $UM \gets getUncertainMetrics(input = FS)$
\If{$isEmpty(UM)$}
\State $IS \gets FS$
\State $LFT \gets getLastTimeStep(FS)$
\Else
\State $idle(until = LFT)$ \Comment{wait for next interval}
\State $FM \gets fetchMetrics(metrics = UM)$
\State $IS \gets temporalMerge(FM, FS)$
\EndIf
\EndWhile
\end{algorithmic}
\end{algorithm}

In summary, the aforementioned pipeline allows for targeted sampling of individual metrics and makes use of probabilistic forecasts if sufficient knowledge is available.

\section{Preliminary Results}
\label{sec:results}

In this section, we examine a prototypical implementation of our monitoring pipeline, called \emph{AM-PF}, obtain preliminary experimental results, and discuss our findings in detail.

\subsection{Data Acquisition}

\begin{figure}
    \centering
    \includegraphics[width=.8\columnwidth]{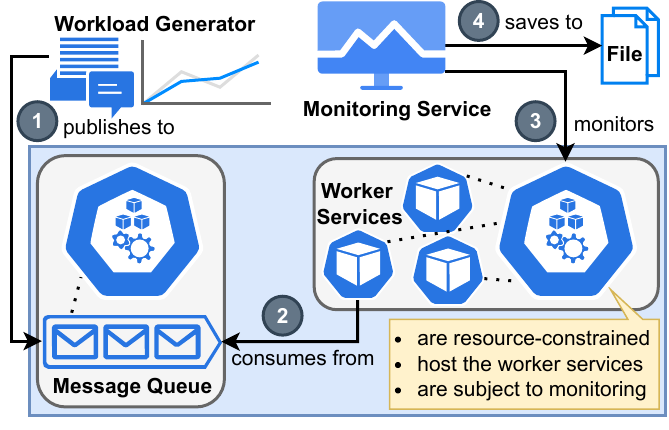}
    \caption{Illustration of our experiment setup for the data acquisition.}
    \label{fig:approach_setup}
\end{figure}

A dataset is required for the evaluation of our proposed framework.
For the sake of highlighting the resource overhead of frequent monitoring, we consider a sampling frequency of 1 second.
Our general goal is to obtain realistic metrics of a node with an underlying predictive pattern, which requires a corresponding workload.
The base workload is created by means of employing 10 python services that are running on the respective node.
Each service consumes messages from an external RabbitMQ message queue, and upon message receipt, several processes are in parallel running basic operations in order to stress relevant resources such as CPU, memory, etc.
These operations are by design dependent on the message rate so that varying patterns can be simulated.
The metrics of the host machine are eventually gathered via Prometheus queries and saved to file. 
As a source for publishing to the message queue, we employ the IoT Vehicles experiment dataset created and published in~\cite{GeldenhuysPSTK22}, which reports amounts of moving cars at intervals of 1 second. 
At each point in time, the corresponding vehicle amount is read and used for publishing the same amount of messages to the message queue.
The whole procedure is illustrated in~\autoref{fig:approach_setup}.

\subsection{Prototype Pipeline}

We implement a prototype of our envisioned pipeline.
First, the values of each variable of the potentially multivariate time series are min-max normalized along the time dimension, where the boundaries are determined from the respective training data and used during inference as well.
We further use as additional features cyclical encodings of \emph{SecondOfMinute} and \emph{MonthOfYear}, as well as custom encodings for 1) \emph{MinuteOfDay} and 2) separating work days and the weekend.

\begin{table}
\small
\centering
\caption{Model Hyperoptimization}
\begin{tabular}[t]{cP{0.55\linewidth}}
    \toprule
    \multicolumn{2}{c}{\emph{Configuration and Search Space}}\\
    \toprule
    Learning rate & 0.001\\
    \#Epochs & max. 20\\
    \midrule
    Input dimension & \{\textbf{300}, 600\}\\
    Output dimension & \{\textbf{300}, 600\}\\
    Hidden dimension & \{25, \textbf{75}\}\\
    Batch size & \{\textbf{256}, 512\}\\
    Dropout rate & \{5\%, 10\%, \textbf{20\%}\}\\
    \bottomrule
\end{tabular}
\label{tbl:hyperopt}
\end{table}

The neural network is implemented using the Darts\footnote{\url{https://unit8co.github.io/darts/}, accessed: October 2022} library, composed of two stacked LSTM layers with a dropout layer in-between and a final linear layer at the end, trained by employing a quantile regression loss, and evaluated with respect to the $\rho$-risk metric~\cite{SeegerSF16}.
Optimized model hyperparameters are found via a hyperparameter tuning approach based on grid search. 
The investigated values are listed in~\autoref{tbl:hyperopt}, with the best ones found highlighted in bold.
The model is subsequently fully-trained with the best found hyperparameters using early stopping, where the training of the model is stopped if the loss of the model on the validation data does not decrease more than 0.001 after 5 steps.
The goodness of the hereby received trained model is further verified via K-fold cross-validation.

\subsection{Baseline}

We compare our method against the AM-DR framework~\cite{FathyBT18} and make use of its publicly available implementation\footnote{\url{https://github.com/YasminFathy/AMDRIoT}, accessed: October 2022}.
In short, this framework employs models both on a source node and respective sink node, such that the former has to transmit only its immediate sensed values that
deviate significantly from the predicted values.
We investigate different maximum error thresholds for this method, while all other parameters are set as reported in the original publication.
Though not directly comparable to our approach due to a different take on the problem, it is insightful to observe the general saving potential on data transmissions as well as implications for metric reconstruction accuracy at sink nodes.

\subsection {Evaluation Setup}

We use the previously acquired data to evaluate both approaches by indexing the respective data file by time and extracting relevant input sequences as arguments to the models, i.e., no additional services are involved in this simplified scenario which favors a detailed model comparison.

For both methods, we evaluate thresholds from 0.005 to 0.05, with a step size of 0.0025. 
For AM-DR, this threshold translates to the maximum error allowed, whereas, for our approach, the threshold marks the largest standard deviation tolerable. 
Due to the different meanings of the threshold in both methods, its configuration is not directly comparable but still allows for insights and the derivation of recommendations.

In terms of evaluation metrics, we are interested in the percentage of transmitted data given various threshold configurations as well as the hereby inflicted implications on the prediction accuracy and metric reconstruction at the sink.

\subsection{Results}

At any point in time, the confidence of the model in its predictions is evaluated and compared against a threshold.
Naturally, the choice of the threshold has an impact on data transmission rates and prediction accuracies, as it controls the monitoring frequency as well as the amount of transmitted data. 
In our experiments, we for instance observe that with a more conservative threshold of 0.0225, our framework is fetching real metrics more frequently when compared to a used opportunistic threshold of 0.0475. 
While the latter requires less fetching and hence relieves network bandwidth usage, it leads to a less accurate observable result.
This manifests itself in higher accumulated Mean Squared Error (MSE) values for higher thresholds (up to 5\% increase in our example), which can also greatly vary across metrics.

\begin{figure}
    \centering
    \includegraphics[width=\columnwidth]{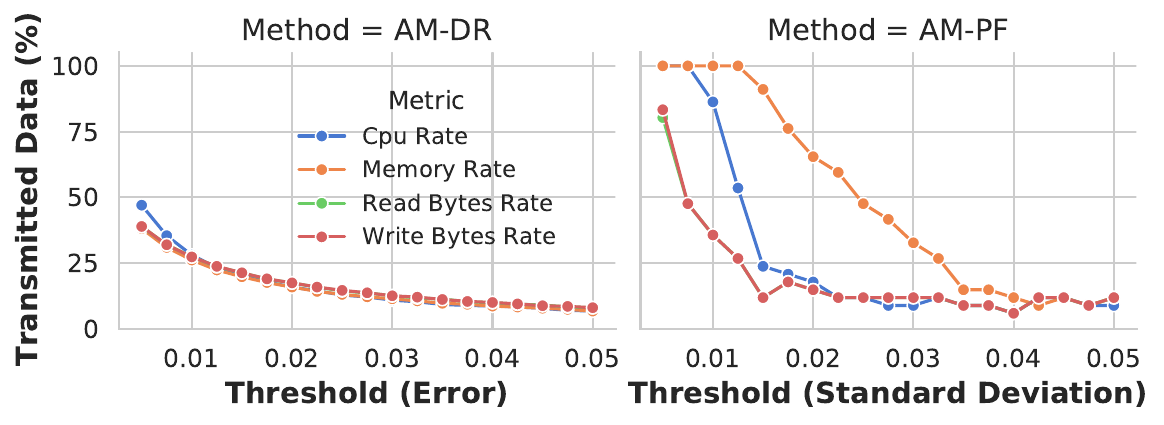}
    \caption{Comparison of frameworks with regard to transmitted data.
    Though not directly comparable, it can be observed how individual metrics are handled with varying levels of sensitivity, based on their inherent characteristics.}
    \label{fig:evaluation_comparison_transmission}
\end{figure}

The percentage of transmitted data in relation to used thresholds is illustrated in~\autoref{fig:evaluation_comparison_transmission}.
For both methods (the results are not directly comparable), we observe that less data is transmitted with increasing thresholds. 
Worth mentioning is that for AM-DR, the reduction is comparably smooth and also similar across metrics, whereas, for our framework, we observe that the decline is initially very steady and also dependent on the specific metric, which indicates that individual metric characteristics are taken into consideration, which favors a more metric-specific adaptation of sampling frequencies.

\begin{figure}
    \centering
    \includegraphics[width=\columnwidth]{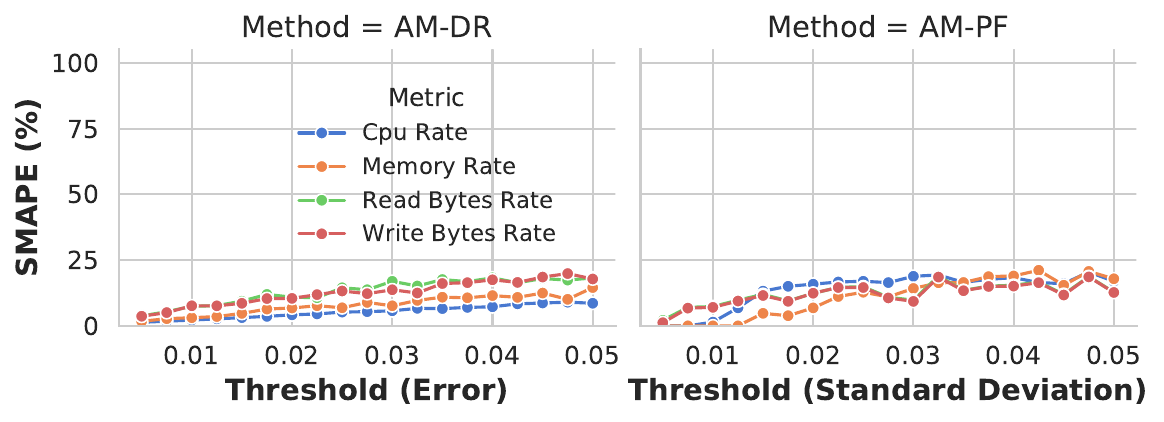}
    \caption{Comparison of frameworks with regard to reconstruction accuracy.
    The prediction discrepancy is measured using the SMAPE metric.
    It can be observed that both methods expose a different convergence behavior.}
    \label{fig:evaluation_comparison_recon_acc}
\end{figure}

Lastly, we also present the prediction errors in relation to used thresholds in~\autoref{fig:evaluation_comparison_recon_acc}.
Here, we use the Symmetric Mean Absolute Percentage Error (SMAPE) since it allows for an easy-to-interpret percentage error as well as comparison across metrics. 
As expected, the SMAPE is increasing for both methods with rising threshold values, since fewer actual metrics are considered and hence the accuracy of inferred metrics is affected.
Noticeably, the SMAPE values of AM-DR are steadily increasing and tend to fan out over time, whereas the SMAPE values of AM-PF are more volatile in the beginning and tend to converge toward a common value. 

Summarizing our results, we find that a suitable configuration of AM-PF yields a similar performance as related methods, while requiring no changes to target edge devices.

\subsection{Discussion}

Our findings demonstrate that probabilistic forecasts can be used to motivate a sampling-based monitoring approach that is able to adapt to data stream changes and reflect on its own recommendations.
By tuning a configurable threshold parameter, the amount of transmitted data as well as the hereby possible metric accuracy can be controlled and balanced. 

For the application of our adaptive monitoring framework in real-world scenarios, we deem the following things important.
First, it is required that target edge nodes are offering metric endpoints which can be scraped -- though a reasonable assumption, some edge nodes might need adaptations if they previously only followed a push-based messaging principle. 
Another aspect to keep in mind is the configuration of the forecasting horizon -- it is advisable to choose a horizon that allows reacting flexibly to any changes that may occur on the edge node.
Furthermore, a resource shortage on the edge node and a need for efficient resource usage must be given, otherwise, the employment of related methods like AM-DR (with their respective overhead) might be more conceivable.
Lastly, while having designed the framework for use in resource-constrained edge environments, we assume that sufficient resources are available at the sink node in the cloud for the model training.

\section{Conclusion}
\label{sec:conclusion}

The primary goal of this work is to demonstrate the applicability of probabilistic time series forecasting for adaptive monitoring in edge computing environments.
To this end, we envision a system that realizes adaptive monitoring in a sampling-based fashion using the aforementioned technique, such that overall network usage is reduced and constrained resources are relieved. 
Towards this goal, we implemented an approach for adaptive sampling based on probabilistic forecasts, and evaluated it in experiments and against a method from related work.
We find that our solution is generally able to reduce the amount of data transmission and furthermore provides means of automating the retraining process.

In the future, we plan to leverage our findings and make use of the proposed methods in the context of recent research on carbon-aware computing in edge environments.

\section*{Acknowledgments}
This work has been supported through grants by the German Federal Ministry of Education and Research (BMBF) as BIFOLD (funding mark 01IS18025A) and by the Deutsche Forschungsgemeinschaft (DFG, German Research Foundation) as FONDA (Project 414984028, SFB 1404).

\bibliographystyle{IEEEtran}
\bibliography{bib}

\end{document}